\documentstyle[epsfig]{mn}
\input{epsf}

\def\zycki{$\dot{\rm Z}$ycki}

\begin{document}

\title[The X--ray Spectrum of Cyg X--2]
{The X--ray Spectrum of Cyg X--2}
\author[C. Done, P.T. \zycki\ and D. A. Smith]
{C. Done$^1$, P. T. \zycki$^2$\ and D. A. Smith$^3$\\
$^1$Department of Physics, University of Durham, South Road, Durham, DH1 3LE, 
UK\\
$^2$N. Copernicus Astronomical Center, Bartycka 18, 00-716, Warsaw, Poland\\
$^3$University of Maryland, College Park, Maryland, MD20742, USA\\
}
\maketitle
 
\begin{abstract}

The spectra of disc accreting neutron stars generally show complex
curvature, and individual components from the disc, boundary layer and
neutron star surface cannot be uniquely identified. Here we show that
much of the confusion over the spectral form derives from inadequate 
approximations for comptonization and for the iron line.
There is an intrinsic low energy cutoff in comptonised spectra at 
the seed photon energy. It is very important to model this correctly
in neutron star systems as these have
expected seed photon temperatures (from either the neutron star
surface, inner disc or self--absorbed cyclotron) of
$\approx 1$ keV, clearly within the observed X--ray energy band.
There is also reflected continuum emission which must accompany the
observed iron line, which distorts the higher energy spectrum. 
We illustrate these points by a reanalysis of the {\sl GINGA} 
spectra of Cyg X--2 at all points along its Z track, and show that the
spectrum can be well fit by models in which the low energy spectrum is
dominated by the disc, while the higher energy spectrum is dominated
by comptonised emission from the boundary layer, together with its
reflected spectrum from a relativistically smeared, ionised disc.

\end{abstract}
 
 
\section{INTRODUCTION}

Disc accreting neutron star systems are amongst some of the brightest
X--ray sources in the sky. They can be split into two broad types
namely atolls and Z sources (Hasinger \& Van der Klis
1989). Observationally these differ in that they show distinct types
of spectral variability and timing behaviour, which link more
fundamentally to a difference in the accretion rate and probably also
in magnetic field (Hasinger \& Van der Klis
1989). The atolls have an accretion rate which is
typically below $\sim 10$\% of the Eddington limit,
and have weak magnetic fields
while the Z sources accrete at $50-150$ \% 
of Eddington, and probably have stronger magnetic fields 
(e.g. the review by Van der Klis 1995). Both
types generally show 2--20 keV spectra which exhibit pronounced
curvature, probably due to Compton scattering of seed photons by
electrons with temperature $kT_e\sim 5$ keV (e.g. White, Stella \&
Parmar 1988), except at low luminosity states which are only seen in
the atoll systems, where the
spectrum makes a transition to a power law (e.g. Mitsuda et al., 1989;
van Paradijs \& van der Klis 1994; Barret et al., 2000).

Here we concentrate on the X--ray spectra of the Z sources.  A full
description of the accretion flow for these sources
should take into
account the interaction of the disc with the stellar surface,
including the effects of radiation pressure and magnetic field in a
relativistic potential. This is likely to be extremely complex, as
shown by initial hydrodynamic (Popham \& Sunyaev 2001; Inogamov \&
Sunyaev 1999) and magnetohydrodynamic simulations (Miller \& Stone
1997). The boundary layer/inner disc emission can then illuminate the
rest of the disc giving rise to a Compton reflected spectrum and
associated iron K$\alpha$ line from the heated, ionised, upper layer
of the disc (see e.g. the review by Fabian et al., 2000, with specific
application to Z sources by Brandt \& Matt 1994). Given this complexity
it is perhaps unsurprising that the observed spectra from
Z sources (and atolls) have proved difficult to fit 
(e.g. Mitsuda et al. 1984; White, Stella \& Parmar 1988). This 
led to the widespread use of colour--colour diagrams as opposed to
true spectral fitting. Here we show that previous difficulties were
mainly caused by using poor approximations for the comptonised and
reflected emission components. We include relativistic smearing
(Fabian et al., 1989) of the disc emission as the source of the line
broadening (which is otherwise very difficult to explain) and derive
excellent fits to the GINGA spectra of Cyg X--2 at all points on its Z
track.

\section{X--ray emission models}

X--rays can arise from a variety of processes from accretion onto a
low magnetic field neutron star.  There can be components from the
neutron star surface, the boundary layer and the X--ray illuminated
disc, and these should be subject to general and special relativistic
effects.  The thermal emission from the neutron star surface is
probably buried beneath a very optically thick boundary layer at the
high mass accretion rates seen in the Z sources (Popham
\& Sunyaev 2001), leaving only the disc and boundary layer as the
major emission components.  The energy release in the boundary layer 
heats the material, which then cools predominantly by comptonization
of any seed photons. The seed photons can either be from 
the disc, or from the neutron star surface, or from 
self--absorbed cyclo/synchrotron photons from the magnetic field 
(Lamb 1989; Psaltis, Lamb \& Miller 1995 but see Wardzinski
\& Zdziarski 1999). The resulting spectrum is then either clearly
comptonised, or quasi--blackbody if the optical depth is
large enough to mostly thermalise the emission (saturated comptonization).
Further Compton scattering, including bulk motion
effects, can take place if the accreting material has a substantial
scale height, so that the emission from the boundary layer has to
travel through this infalling material before it can escape to the
observer (Lamb 1989; Psaltis et al., 1995). 

The boundary layer can directly illuminate some fraction of the disc. 
Part of the illuminating
flux is reflected (Smale et al 1993; Brandt \& Matt 1994), while the
rest is thermalised, enhancing the disc emission (e.g. Czerny, Czerny
\& Grindley 1986).  This heats the upper layers of the disc to the
Compton temperature of around 1--2 keV. This exceeds the
virial temperature at distances $\ge 3-6\times
10^{10}$ cm for a $1.4M_\odot$ neutron star,
so the material escapes as a wind forming a corona
around the outer accretion disc (Begelman, McKee
\& Shields 1983). This accretion disc corona (ADC) is especially important in
determining the observational appearance of the highly inclined
systems, where the intrinsic, central source is hidden by the disc but
can be seen via scattering in this tenuous material above and below
the disc (e.g. White, Nagase \& Parmar 1995). The photo--ionised ADC
emits line and continuum radiation, and can also scatter some of the
boundary layer/inner disc emission back onto the disc, increasing the
illumination (e.g. Vrtilek, Soker \& Raymond 1993).  Even at somewhat
smaller radii the X--ray illumination heats the upper layers of the
discs, so it has an appreciably larger scale height than an
unilluminated disc, although the stronger gravity means that it cannot
escape to form a wind.  The emission from this photo--ionised, heated
layer consists of recombination features (line and continuum
radiation) together with the reflection of some fraction of the
incident X--rays from the disc (Kallman \& White 1989; Vrtilek, Soker
\& Raymond 1993; Ko \& Kallman 1994).

\section{Spectral Model}

From the discussion above, the 
spectrum should consist of components from the disc and boundary
layer, together with reflection from the disc 
and scattering/emission/absorption from the ADC. Generally progress is
made by estimating which components dominate so as to form a
simplified model which nonetheless describes most of the physical
processes.  This is where the problems arise as there is no consensus
on which effects dominate. 

One approach to modelling the spectra (generally called the `Eastern
model' in the literature) assumes that an optically thick disc
dominates at low energies, while a Comptonised boundary layer
dominates at higher energies. This was developed by Mitsuda et al.,
(1984; 1989), and fits the data well with the addition of a {\it broad}
iron K$\alpha$ line (White, Peacock \& Taylor 1985; Hirano et al.,
1987; Asai et al., 2000).

However, spectral fitting is inherently non--unique, and another 
approach is to assume that the majority of the
emission at both high and low energies comes from a comptonised disc,
while the boundary layer emits as a quasi--blackbody in the middle of
this energy range (White et al., 1988). Such models 
(generally termed `Western' models) 
can fit the data equally well, but again require the inclusion of a
broad iron line component (White et al., 1988; Mitsuda et al., 1989;
Hasinger et al., 1990).

Here, we follow the Eastern approach, as it seems most likely that the 
boundary layer is strongly Comptonised (Popham \& Sunyaev 2001). We
assume that the low energy X--ray continuum is dominated by the
disc, but the disc spectrum is not well known especially at accretion rates
close to Eddington where radiation pressure must be important. The
radial temperature distribution depends on whether the viscosity
is proportional to the gas pressure or the total (gas plus radiation)
pressure (Stella \& Rosner 1984). 
Electron scattering in the atmosphere is also expected to be
important, leading to weak Comptonization of the disc emission
(Shakura \& Sunyaev 1973), and the intrinsic disc spectrum
should be further broadened by relativistic effects (Ebisawa et
al., 1991). Detailed calculations of these distortions show that
with a restricted bandpass, where only the high energy tail of the
disc spectrum is seen, then the shape of the spectrum can be
well approximated by a simple multi--colour disc model with a colour
temperature correction (Ebisawa et al., 1991; Shimura \& Takahara
1995; Merloni et al. 2000). 
Thus we approximate the disc spectrum by the {\tt diskbb} model in
XSPEC, but also test that this assumption does not strongly bias our
results by using other disc spectral models. 

We assume that the seed photons for the comptonization in the boundary
layer come from the neutron star surface as well as the inner disc, so
they can be at higher energies than the observed disc emission. The
typical temperatures expected from close to the neutron star at high
mass accretion rates are $\approx 1$~keV, i.e. within the observed
X--ray bandpass. Simple analytic approximations (e.g. the {\tt compst}
model or a cutoff power law) for the Compton scattered spectrum are
only valid where the seed photon energy is very much lower than that
of the hot electrons (e.g. Hua \& Titarchuk 1995).  Explicit seed
photon comptonization models are required (such as the {\tt comptt}
model in XSPEC), and these give spectra with significantly different
curvature close to the seed photon energies.  Such models are starting
to be used to fit broad band (generally SAX) data from Z and atoll
systems (Guainazzi et al., 1998; Piraino et al 1999; in't Zand et al.,
1999; di Salvo et al., 2000; 2001). However, the 
assumption in {\tt comptt} is
that the seed photons have a Wien distribution. Here instead we
assume that the seed photons have a blackbody distribition
and use an analytic
approximation to the Comptonisation 
based on the solution of Kompaneets equation (Zdziarski, Johnson \&
Magdziarz 1996), as described in \zycki, Done \& Smith (1999).

This comptonising cloud around the neutron star can illuminate the
accretion disc, ionising it and 
producing a characteristic reflected component and
iron K$\alpha$ line.  This was first detected in Active Galactic
Nuclei, and extensively studied both in supermassive and in 
in Galactic Black Holes (see e.g. Fabian
et al., 2000 and references therein). However, its application to
spectra from neutron star systems is much less widespread. The atoll
systems at low mass accretion rate have power law spectra in the 2--20
keV range, and reflection is sometimes included to fit these data
(Yoshida et al., 1993; Piraino et al 1999; Barret et al., 2000). 
However, the curving spectra from higher accretion rate atolls and Z
sources have not been fit with reflection models
despite observational (Smale et al., 1993) and
theoretical (Brandt \& Matt 1994) motivation. 

Another issue of importance in modelling the reflected spectrum is 
relativistic smearing. 
The disc extends down close to the
surface of the neutron star, so reflection from the {\em inner} disc with 
relativistic distortions on both
the line and reflected continuum should be important (Fabian et al
1989). This has been seen in AGN (Tanaka et al., 1995) and Galactic
Black Hole Candidates (\zycki, Done \& Smith 1997; 1998; 1999; Done \&
\zycki\ 1999) but again has not been studied in 
high mass accretion rate neutron star systems. Instead the observed
strong broad line at 6--7 keV (White et al., 1985; Hirano et al.,
1987; Asai et al. 2000) is generally interpreted as arising from the
ADC and illuminated {\em outer} disc (Kallman \& White 1989; Raymond 1993),
although it is very difficult to make both the observed line intensity and 
width from such a corona (see e.g. the discussion in Smale et al.,
1993).

We calculate the reflected emission from the {\tt thcomp} continuum as
described in \zycki\ et al., (1999). This calculates the self
consistent line emission as well as the angle dependent reflected
continuum for any ionization state and iron abundance (all the other
elements are assumed to have solar abundance: Morrison \& McCammon
1983). 
Relativistic smearing (including light bending) is then
applied to this total (line plus continuum) spectrum (Fabian et al.,
1989). The resultant
spectrum depends on the inclination, which 
we fix at $60^\circ$ (Orosz \& Kuulkers 1999).

We assume that emission/absorption/scattering from the ADC above the
outer disc is negligible. This is probably not the case for the low
energy spectrum, as line emission is seen at $\sim 1$ keV in several
LMXB's (e.g. Bautista et al., 1998 and references therein)
including Cyg X--2 (Kuulkers et al., 1997). However, here
we are concerned with the 2--20 keV spectrum, and the iron K
line associated with the ADC is probably small as the K/L line
ratio is low for the steep spectra characteristic of the Z
sources (Kallman 1995).

\section{OBSERVATIONS}

\subsection{GINGA data}

\begin{figure}
\begin{tabular}{c}
\epsfig{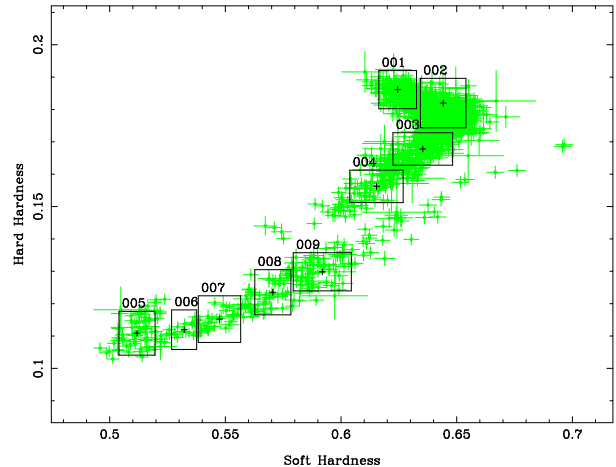}
\end{tabular}
\caption{A colour -- colour plot showing the Z track of Cyg X--2 in
the June 1988 data. The overlaid boxes show the data used to make the
individual spectra}
\label{fig:col_col}
\end{figure}

\begin{table}
\caption{GINGA observations of Cyg X--2 in June 1988}
\begin{tabular}{ccc}
\hline 
& UT Start & UT End \\
\hline
& 88-Jun-10 12:35:00 & 88-Jun-11 05:25:54 \\
& 88-Jun-11 13:30:00 & 88-Jun-12 22:30:26 \\
& 88-Jun-13 13:15:00 & 88-Jun-14 04:00:34 \\
\end{tabular}
\end{table}
 
\begin{figure*}
\begin{tabular}{cc}
{\epsfig{file=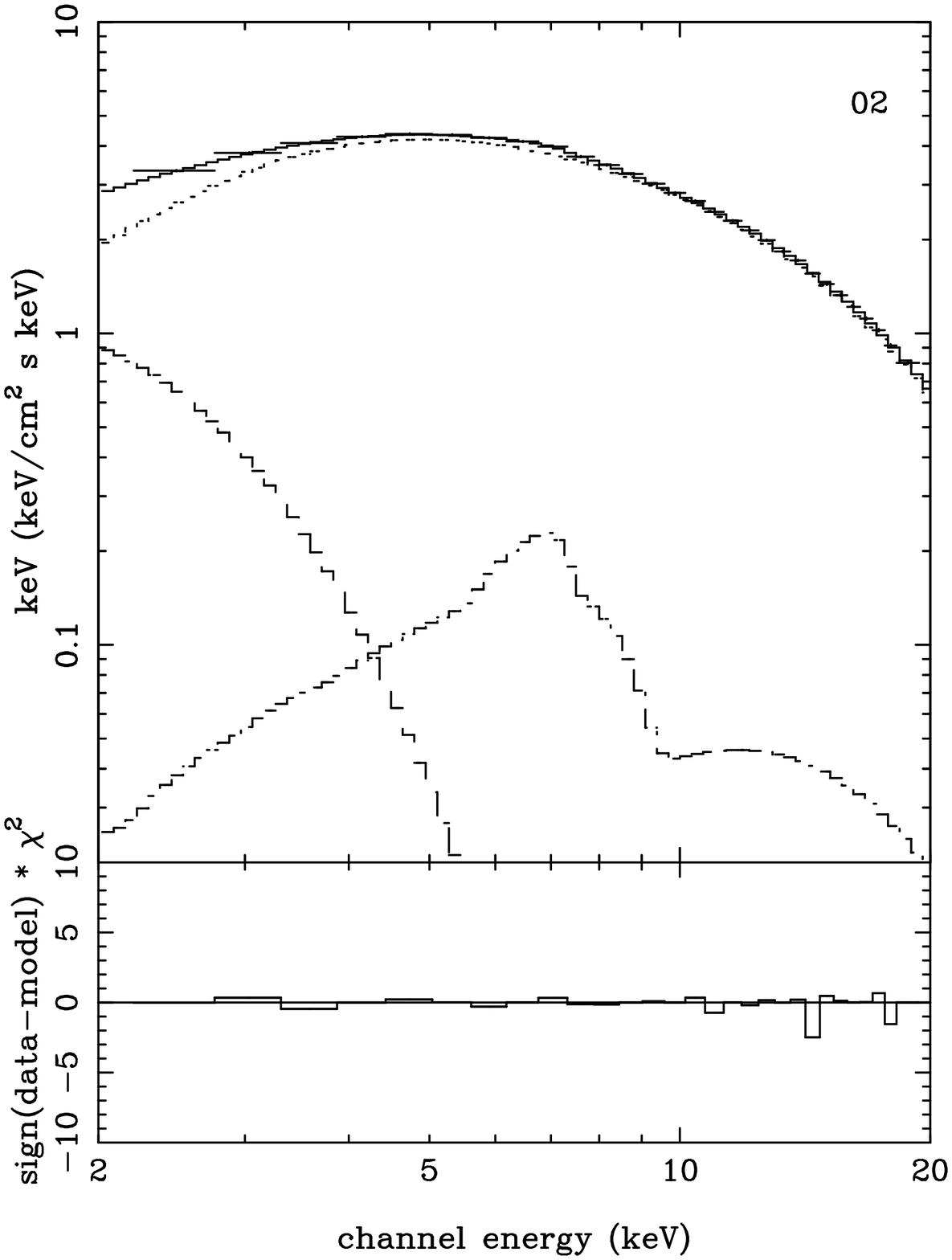, width=0.45\textwidth}}
&
{\epsfig{file=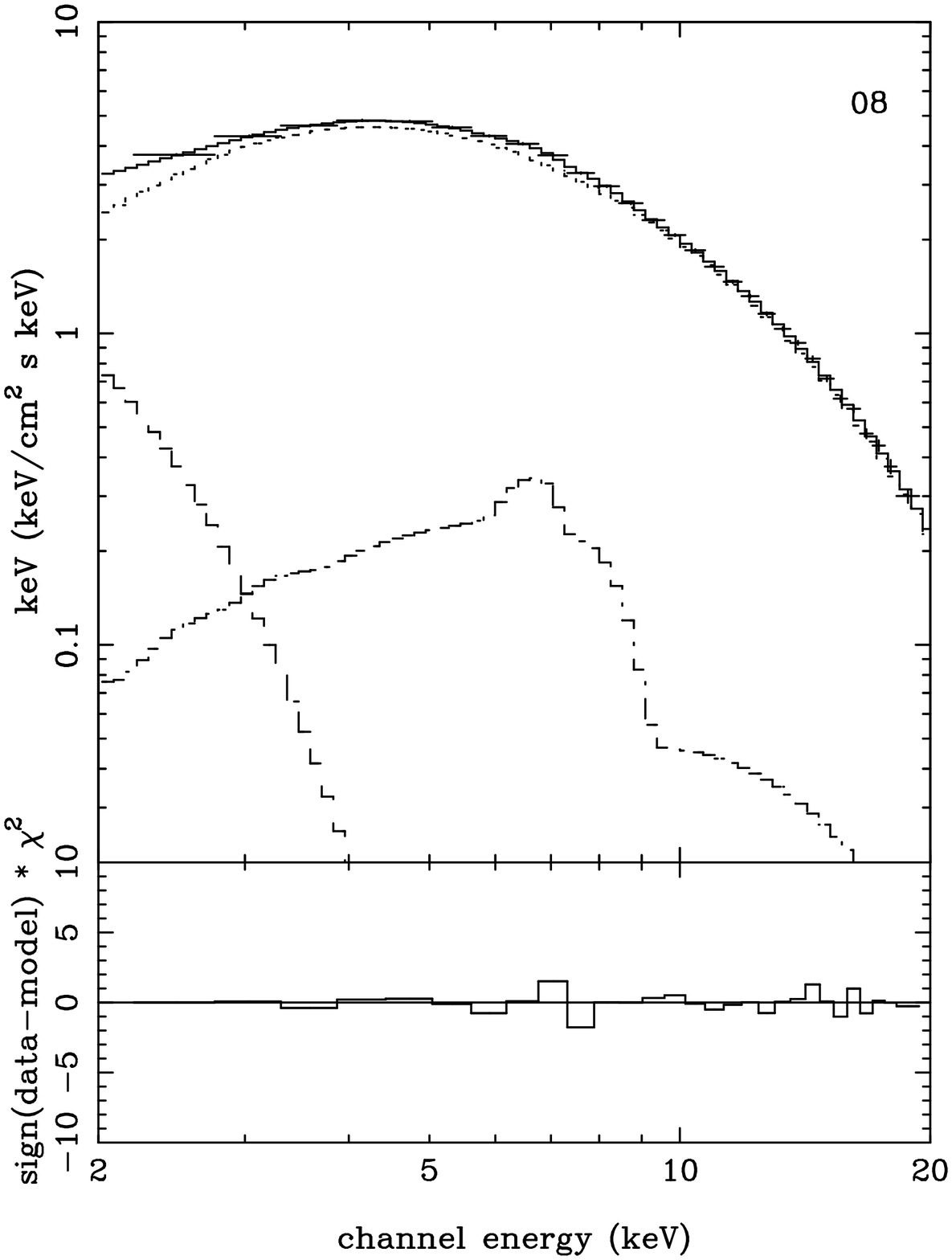, width=0.45\textwidth}}
\end{tabular}
\caption{The two panels show spectral fits and residuals to 
02 (horizontal branch: left hand panel)
and 08 (flaring branch: right hand panel) 
spectra using the models detailed in Table \ref{tab:dcref}.
The low energy component shown by the dashed line is the disc
blackbody, the dotted line is the comptonised continuum spectrum, the
dashed--dotted line is its reflection from the disc while the solid
line shows the total spectrum. 
The softening of the high energy emission is clearly the major
difference between the horizontal and flaring branch spectra.}
\label{fig:dcref}
\end{figure*}

We use the Cyg X--2 {\sl Ginga\/} data, 
re--extracting them from the original FRF's using the {\sl
Ginga\/} data reduction software at Leicester. Observations were
performed in MPC2, MPC3 and PC modes (see Hasinger et al., 1990);
however, in this paper, we use only the MPC2 mode data in order to
perform a detailed spectral analysis. Data were rejected from periods
when the SUD rate ({\it i.e.\/} the LAC count rate above 24~keV)
exceeded 8~counts/sec/detector, the magnetic cut-off rigidity was
outside the range 7--20~GeV/c, the angle between the LAC pointing
direction and the Earth's horizon was outside the range
6--120~degrees, and the SOL2 rate (of the solid-state electron
detector on board {\sl Ginga\/}). Unfortunately, the intensity of
Cyg X--2 during the observations was such that the register storing
the detected count rate overflowed occasionally before telemetry. This
affected only data accumulated at medium bit-rate, and could be
corrected after examining each PHA spectra. The data were corrected
for deadtime using the following formulae:
$$
\lambda = \frac{\mu}{1 - \mu_{\rm DET}\times\tau}, \,
\sigma^{2}_{\lambda}  = \frac{\sigma^{2}_{\mu}}{(1 - 
\mu_{\rm DET}\times\tau)^{2}}
$$
where, $\lambda$ and $\sigma^{2}_{\lambda}$ are the true count rate
and variance, $\mu$ and $\sigma^{2}_{\mu}$ are the observed count rate
and variance, $\mu_{\rm DET}$ is the observed count rate for one
detector integrated over all PHA channels, and $\tau$ is the deadtime
($ = 1.8 \times 10^{-4}$ seconds). Background subtraction was
performed using off-source data taken in a contemporary 3--4 month
period, and correcting for the 37-day periodicity of the {\sl Ginga\/}
orbit ({\it i.e.} the ``universal'' method of Hayashida et al., 1989;
Williams et al., 1992). The spectrum of Cyg X--2 is sufficiently soft
as to not contaminate the SUD count rate, unlike other X-ray binaries
such as Cyg X--1. Finally, we note the the Galactic emission
estimated from a couple of off-source regions near to Cyg X--2 is
negligible.

The background subtracted data were rebinned on a 40~seconds
timescale. The soft hardness (4.5--9.0~keV/1.1--4.5~keV) versus hard
hardness (14.2--9.0~keV/4.5--9.0~keV) ratio for the June and October
datasets are shown in Figure \ref{fig:col_col}. 
These are identical to those presented
by Hasinger et al., 1990. In the Hasinger et al., analysis, spectra were
extracted at various intervals during the observations; to avoid
spectral variations, these spectra were of short exposure. Here we
extract spectra from each box in Figure \ref{fig:col_col}, 
thus enabling us to obtain
much longer exposures {\it but} avoiding any significant spectral
variations and so a much more detailed analysis can be made. For the
remainder of this paper, we refer to the spectra as 01--09, respectively.

\section{SPECTRAL ANALYSIS}

The interstellar column density to Cyg X--2 determined through visual
extinction to the secondary star and the relative size of the X--ray
dust scattering halo is $2.2\times 10^{21}$ cm$^{-2}$ (Predehl \&
Schmitt 1995). This is consistent with the column seen from ROSAT,
BBXRT, ASCA and SAX spectral analysis so we fix this in all the
following fits. 

Results from fitting models based on the ideas developed in Section 3
are given in Table \ref{tab:dcref}. 
We obtain excellent fits to all the individual
spectra (total $\chi^2_\nu=136.8/180$) with optically thick disc
emission and a Comptonised boundary layer together with its reflection from an
ionized disc. The spectral fits for two different positions on the
Z track (02 and 08) are shown in Figure \ref{fig:dcref}. 

Previous papers have often modelled the iron spectral features using a
broad line. Table \ref{tab:dcgau} 
shows the results obtained by replacing the more
physically motivated reflected spectrum with an {\it ad hoc} broad
iron line. The line energy and intrinsic
width are constrained to be between 6.4-7 keV, and less than
$\sigma\le 0.42$ (corresponding to a FWHM$\le 1$ keV),
respectively, for stability. The fit is adequate (although the model fails to
describe the spectra on the normal branch - flaring branch transition)
with $\chi^2_\nu=211.9/180$. An F test shows that 
reflection is 99.98 per cent more probable to be the correct 
description of the spectral features than a broad Gaussian iron line.

\begin{figure}
\begin{tabular}{c}
\epsfig{file=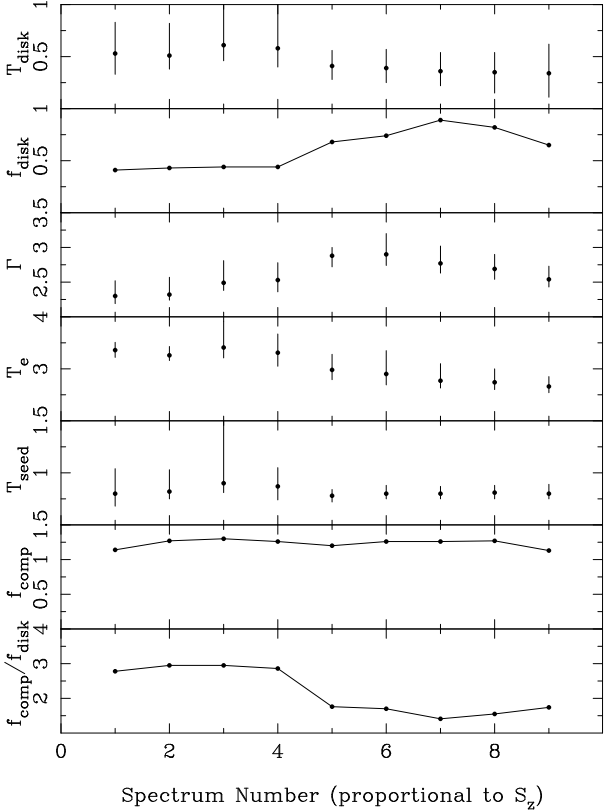, width=8.truecm}
\end{tabular}
\caption{The derived continuum parameters for the spectral fits
to the Eastern model with full boundary layer comptonization and
reflection ({\tt diskbb+thcomp+reflection}) given 
in Table \ref{tab:dcref}
as a function of position along the Z track. All temperatures are
given in keV, and all fluxes are in units of $10^{-8}$ ergs cm$^{-2}$
s$^{-1}$. The only statistically significant changes are that the 
disc blackbody flux increases while the comptonised spectrum softens.}
\label{fig:dcref_c}
\end{figure}

Figure \ref{fig:dcref_c} 
shows how the derived continuum parameters evolve along the Z
track using the best fitting reflection model for the spectral
features (Table \ref{tab:dcref}). 
All fluxes are derived by integrating the relevant
unabsorbed model spectrum over 0.01--100 keV. The most significant
change in the spectrum is the factor 2 increase in disc flux between
spectra 4 and 5. However, despite this increase in accretion rate
through the disc, the flux in the boundary layer stays constant
to within 10 per cent, although its spectrum softens 
(seen both by an increase in spectral index and decrease in
electron temperature). Thus the the
ratio of flux dissipated in the boundary layer
also makes an abrupt change between spectra 04 and 05, from 
$\sim 3$ to $\sim 1.6$. 

\begin{figure}
\begin{tabular}{c}
\epsfig{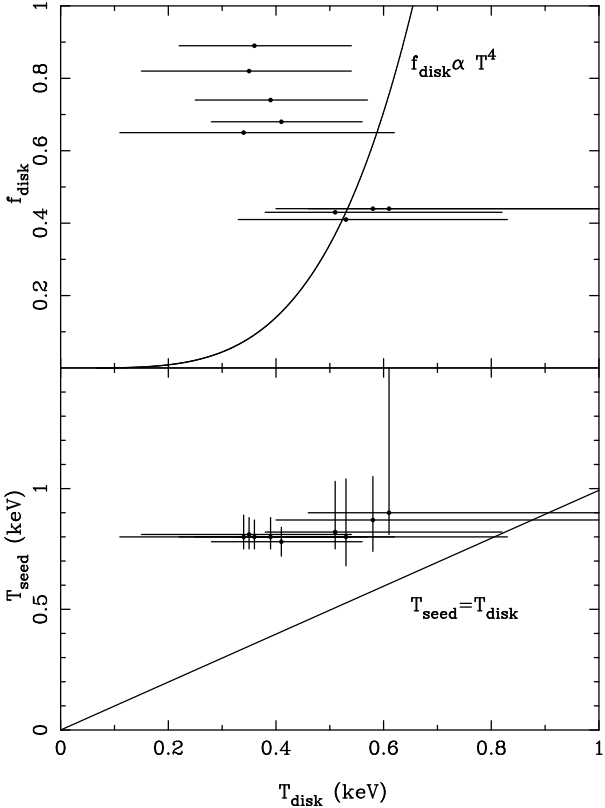}
\end{tabular}
\caption{The soft spectral components from the Eastern model with full 
boundary layer comptonization and reflection 
({\tt diskbb+thcomp+reflection}). The
upper panel shows disc temperature versus disc flux 
from the fits 
in table \ref{tab:dcref}. The disc temperature is consistent with a
constant as the flux varies by a factor 2, which would imply that the
disc inner radius was increasing. However, the uncertainties are large
and the data cannot rule out the 
expected flux--temperature dependence for a
constant radius flow, shown by the solid curve. The lower panel shows
the seed photon temperature as a function of the disc temperature. 
The seed photons are plainly significantly hotter than the observed
disc temperature (the solid line shows $T_{seed}=T_{disc}$), showing
that the observed soft component is not the major
source of seed photons for the comptonizing cloud.}
\label{fig:dcref_d}
\end{figure}

Figure \ref{fig:dcref_d} 
shows the evolution of the soft component in more detail.
The disc temperature is clearly consistent with staying constant as
the flux in the soft component increases (upper panel), which would
imply that the optically thick boundary layer extends 
progressively over more and more of the inner disc, so the the observed
inner disc radius increases. However, the uncertainties are large and
cannot exclude the $T_{disc}\propto f_{disc}^{0.25}$ dependence
expected for increasing accretion rate through a constant radius flow.
The lower panel of figure \ref{fig:dcref_d} 
shows that the seed photons for the
compton scattered component have a temperature which is again
consistent with a constant but is significantly higher than the
temperature of the observed soft component from the disc. This implies
that the disc photons we observe are {\it not} the source of seed
photons, and again, the most probable geometry is one where the
optically thick comptonising cloud surrounds most of the neutron star
surface and also perhaps the innermost disc.

\begin{figure}
\begin{tabular}{c}
\epsfig{file=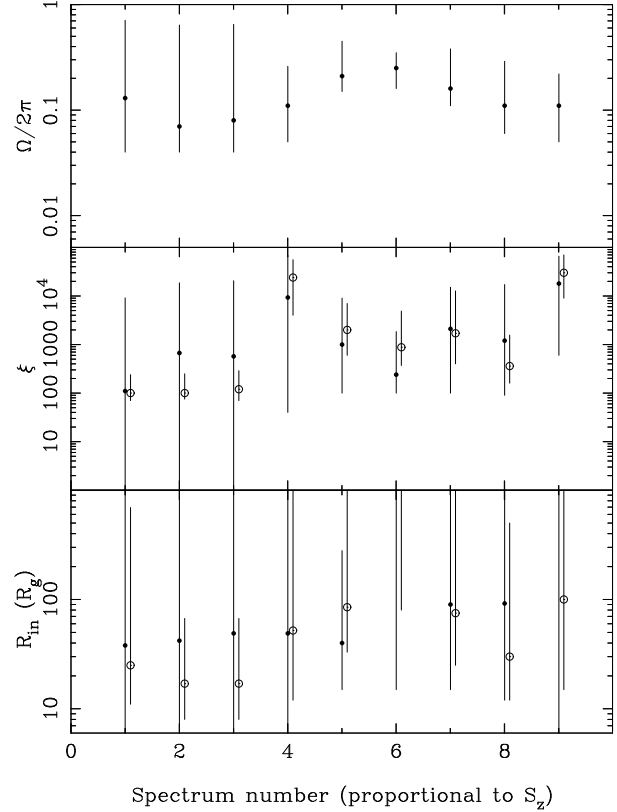, width=8.truecm}
\end{tabular}
\caption{The derived reflected parameters from the Eastern model with full 
boundary layer comptonization and reflection 
({\tt diskbb+thcomp+reflection}) given in 
Table \ref{tab:dcref}
as a function of position along the Z track. All the parameters are
statistically consistent with constant values due to the large
uncertainties, although this masks subtle spectral changes. Assuming
that the solid angle subtended by the reflector is constant then there
are significant changes in the ionization of the 
reflected spectrum. These are shown by the open symbols 
which are offset by 0.1 in spectral number.}
\label{fig:dcref_ref}
\end{figure}

The reflection parameters are shown in Figure \ref{fig:dcref_ref}. 
These are poorly
determined since reflection is a curving spectral component and so is
difficult to disentangle from the other two curving spectral
components from the disc and comptonised boundary layer.  The spectral
features help, but the relative contrast of these can be dramatically
changed by ionization effects (and by relativistic smearing). All the
parameters are consistent with a constant value ($\Omega/2\pi=0.16$,
$\xi=330$ and $R_{in}= 120 R_g$).  However, the parameters are highly
correlated, and {\it together} they are not consistent with these
constant values. Fixing the reflected fraction 
(as appropriate if the geometry remains the same) 
at the mean value of $\Omega/2\pi=0.16$ requires a
significant change in the ionization parameter, as shown by the offset
open symbols in Figure \ref{fig:dcref_ref}. 
Physically, what is happening is that the
spectral features increase in intensity along the Z track,
so either the solid angle subtended by the reflector increases or
its ionization state increases. Since the solid angle here is fixed,
then the ionization state increases. Alternatively of course, it could
be that the ionization state remains fixed and the geometry changes or
that both change. Unfortunately, the broadening of the spectral
features is still largely unconstrained, these data cannot determine
the inner disc radius from relativistic smearing.

\section{ALTERNATIVE SPECTRAL MODELS}

While this model gives a good description of the observed spectrum,
the underlying continuum form is not unique.  Even within the same
physical picture of disc accretion and a comptonizing boundary layer
there are different approaches to modelling the data. There are
several approximate forms for the Comptonised spectrum. Our model,
which is very similar to the {\tt comptt} model in XSPEC (Titarchuk
1994), includes the spectral curvature expected from having the seed
photons within the observed bandpass. Previous papers have often used
the {\tt compst} model (Sunyaev \& Titarchuck 1980) or cutoff power
law for the comptonised flux, but this is applicable {\it only} where
the seed photons for the Comptonization are at much lower energies
than the observed bandpass. This is plainly not the case for the Z
sources (nor for {\it any} neutron star accreting at more than ~0.1
per cent of Eddington). 

\begin{figure}
\begin{tabular}{c}
\epsfig{file=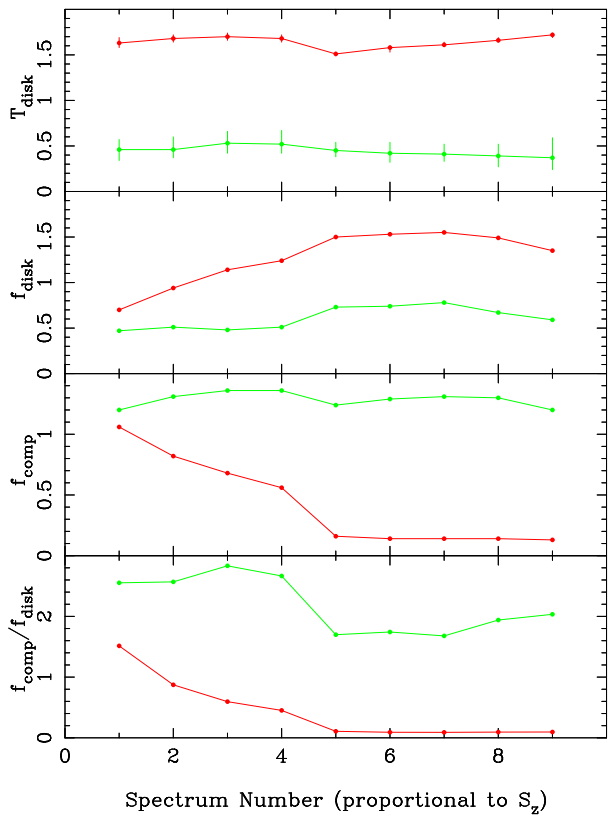, width=8.truecm}
\end{tabular}
\caption{Continuum parameters derived from the Eastern model
assuming approximate boundary layer comptonization ({\tt compst}, dark
grey, Table \ref{tab:dcstgau})
compared to those
derived from models which include the seed photon curvature ({\tt
thcomp}, light grey, Table \ref{tab:dcgau}). Both sets of fits have 
the disk spectrum modelled by {\tt diskbb} and reflection is
approximated by a gaussian line.
The disc temperature is given in
keV, and all fluxes are in units of $10^{-8}$ ergs cm$^{-2}$ s$^{-1}$.
Plainly there is a dramatic difference in the inferred continuum
depending on whether the comptonised spectrum includes the effect of
the seed photons. Given that the seed photon are expected to be within
the observed bandpass then it is clear that approximations to the
comptonised spectrum which do not include this can be severely misleading.}
\label{fig:dcstgau_cont}
\end{figure}

In Table \ref{tab:dcstgau} we fit the phenomenological model of a 
disc blackbody, comptonised continuum without seed
photons and broad iron line) to the data, and obtain very different
results (compare to Table \ref{tab:dcgau} where the comptonised model
includes seed photon curvature).  
Firstly, the fits are much worse using the {\tt compst}
approximation -- typically by 
$\Delta\chi^2\sim 10$ larger in each spectrum 
for one less model parameter (the seed
photon energy).  More importantly, the derived model parameters are
very different.  Figure \ref{fig:dcstgau_cont}
 compares the continuum luminosities derived
from the {\tt compst} model (Table \ref{tab:dcstgau}), 
to those derived from the {\tt
thcomp} model (Table \ref{tab:dcgau}) 
where the curvature in the comptonised
spectrum near the seed photon energy is included.  The disc
temperature is much higher in the {\tt compst} fits, as is its flux,
but its implied radius is much smaller.
 
The mean radius of the inner disc derived from 
the {\tt thcomp} approximation (including seed photon curvature) is
$60^{+13}_{-40}$ km, compared with $9.2\pm 0.2$ km from {\tt compst}
(from the mean normalisation of the {\tt diskbb} components, assuming
that $\cos(inclination)/D_{10}^2\sim 1$ as appropriate for the best
estimates of distance in units of 10 kpc, $D_{10}=0.72$, and
inclination of $60^\circ$: Orosz \& Kuulkers 1999). 
Both are of order
the neutron star radius, especially given the many systematic effects
inherent in the {\tt diskbb} radius (neglect of colour temperature
correction, relativistic effects and stress free inner boundary
condition), but the apparent emitting area 
is suppressed by a factor of 7 by using an 
incorrect spectral form for the comptonised
spectrum. Neglecting the turnover at the seed photon energy means that
the contribution of the comptonised emission is overestimated at low
energies, so the remaining flux to be associated with the disc emission is
underestimated. Previous comparisons the Western and Eastern models 
have cited the very small emitting area of the disc as a problem for
the Eastern models (e.g. Church \& Balucinska--Church 2001), but here
it is clear that the small area can be an artifact of incorrect
spectral modelling of the comptonised component.

\begin{figure}
\begin{tabular}{c}
\epsfig{file=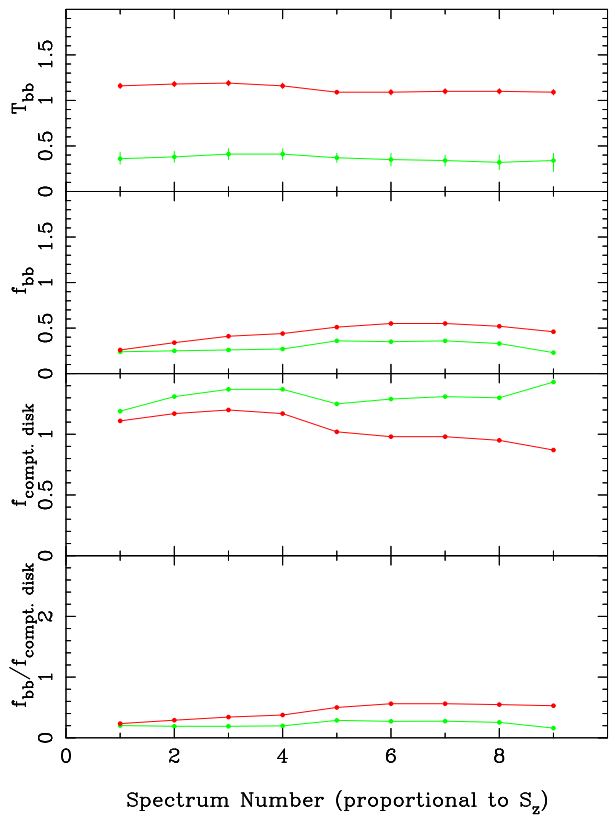, width=8.truecm}
\end{tabular}
\caption{Continuum parameters derived from the Western model
assuming approximate disc comptonization ({\tt compst}, dark
grey, Table \ref{tab:cstbbgau})
compared to those
derived from models which include the seed photon curvature ({\tt
thcomp}, light grey, Table \ref{tab:cbbgau}).
Both sets of fits have the thermalized boundary layer modelled by {\tt bbody}
and reflection is approximated by a gaussian line.
The approximate comptonisation model 
requires a blackbody at $\sim 1$ keV 
as appropriate for the boundary layer, but its luminosity is far below
that of the disc. There is a substantial drop in the derived
'boundary layer' blackbody temperature to $\sim 0.3$ keV 
when the comptonised disc is modelled by the more
physically realistic comptonization 
which includes the curvature expected at
energies close to the seed photons.  The blackbody temperature is given in
keV, and all fluxes are in units of $10^{-8}$ ergs cm$^{-2}$ s$^{-1}$.}
\label{fig:cstbbgau_cont}
\end{figure}

By contrast, changing the description of the disc emission does {\it
not} dramatically change the inferred parameters for fluxes from the
disc and boundary layer components. The {\tt diskbb} disc blackbody
model approximates the disc spectrum neglecting the stress--free inner
boundary condition, relativistic corrections and any energy input from
illumination, and assumes that the disc spectrum at any radius is a
true blackbody (Mitsuda et al 1984). The {\tt diskm} model in XSPEC
(Stella \& Rosner 1984) tackles two of these assumptions, in that it
uses a modified blackbody spectrum to describe the disc spectrum where
the opacity is dominated by electron scattering (Shakura \& Sunyaev
1973), and includes the inner boundary condition. The spectrum is then
somewhat broader than the {\tt diskbb} model, but gives results which
are more or less the same as those for the {\tt diskbb} model in the
2--20 keV range in terms of the total disc flux, and concomitant
comptonised continuum parameters. We fix the {\tt diskm} spectrum
normalisation 
$=\cos(inclination)/D_{10}^2$ at unity 
(Orosz \& Kuulkers 1999), alpha viscosity at
$0.1$ and mass of the neutron star at $1.4M_\odot$.  Using the {\tt
diskm} model with the {\tt thcomp} continuum and broad iron line to
model the spectrum gives an adequate fit (total
$\chi^2_\nu=224.1/180$), but replacing the broad Gaussian line by the
reflected spectrum again gives a significant improvement in the fit
(total $\chi^2_\nu=139.5/180$).

The differences between {\tt diskm} and {\tt diskbb} are quite
subtle. However, if there is substantial comptonization of the disc
emission then the disc spectrum can change dramatically. Such strong
comptonization is sometimes seen in the Galactic Black Holes at high
mass accretion rates (\zycki, Done \& Smith 2001). This is the basis
of the `Western' approach, in which it is the {\it disc} emission
which is strongly comptonised, while the boundary layer is a (more or
less) simple blackbody (White, Stella \& Parmar 1988, see Section
2). Table \ref{tab:cstbbgau} 
shows results from this with the commonly used {\tt
compst} model, together with the phenomenological broad iron line to
model the spectral features. This is a good fit to the spectra (total
$\chi^2_\nu=171.5/189$), and has a blackbody boundary layer
temperature of $\sim 1$ keV, but the ratio of the boundary layer to
disc flux is much lower than that predicted for a non--rotating
neutron star. This was noted by White et al., (1988) and suggested as
evidence for equilibrium (fast!) neutron star spin periods. However,
this result is {\it not} robust. The {\tt compst} model assumes that
the seed photons are far below the observed bandpass, so the spectrum
is a power law below energies corresponding to the electron
temperature. But the seed photons {\it must} be in the observed energy
range, since the assumption here is that it is the disc emission
itself which is being comptonised. Using the {\tt thcomp} model for the
comptonised disc, so that the spectral curvature expected from the
seed photons is included, gives very different results (Table
\ref{tab:cbbgau}). Figure \ref{fig:cstbbgau_cont} shows a comparison of the
derived blackbody boundary layer temperature and flux for these two
different disc comptonization models. Plainly the continuum parameters
vary substantially, and again, we stress that using {\tt
compst} outside of the range where its approximations are valid can
give misleading results. 

\begin{figure*}
\begin{tabular}{c}
\epsfig{file=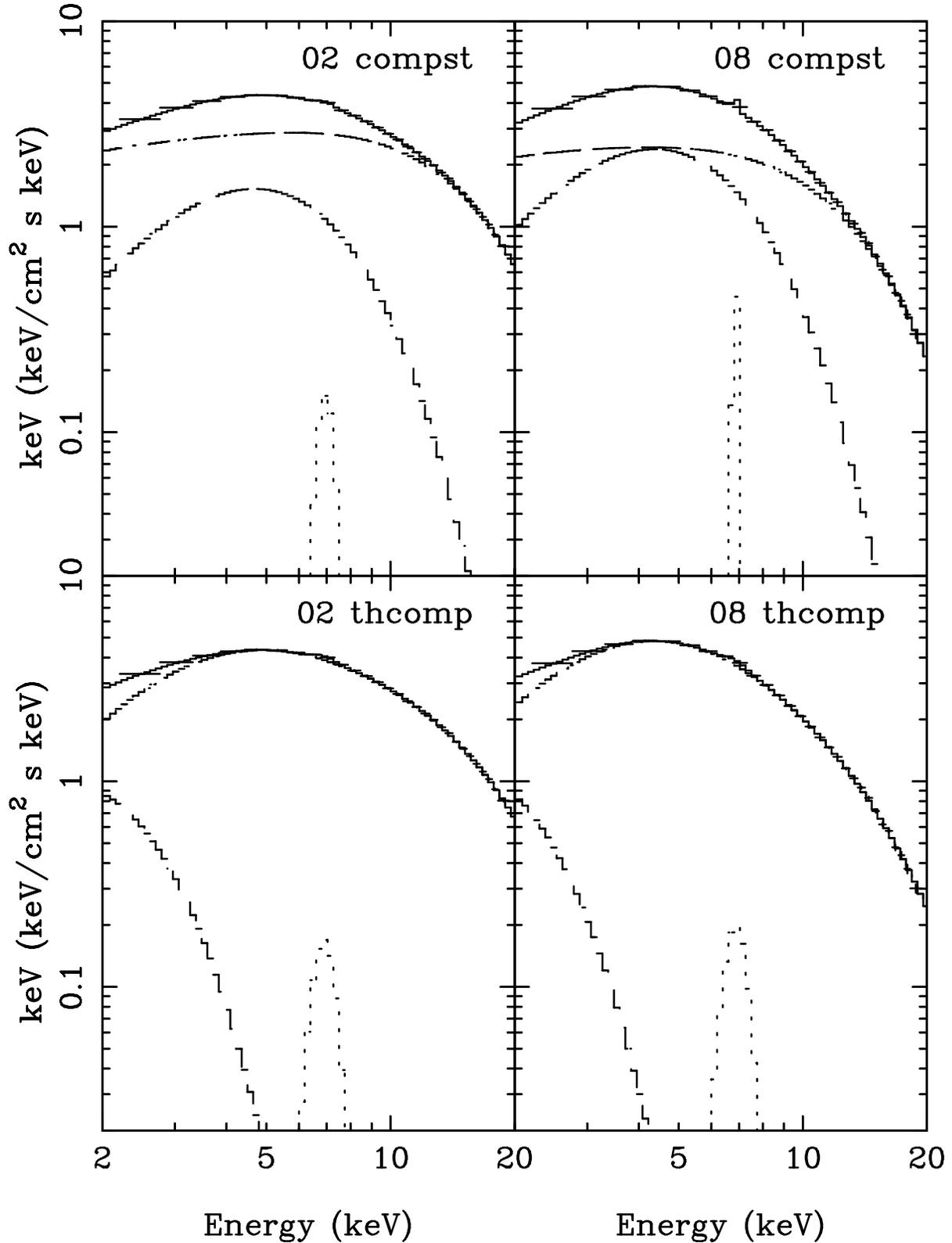, width=0.9\textwidth}
\end{tabular}
\caption{Spectra derived using the Western model, assuming that the
disc emission is strongly comptonised while the boundary layer is
mostly thermalised. The upper panel shows the derived fits to
02 (horizontal branch) and 08 (flaring branch) spectra using the 
{\tt compst} model for the comptonised disc. This does not have the
seed photon curvature included, and the data then require a high
temperature $\sim 1$ keV blackbody component to fit the observed spectra. 
The lower panel shows the same data deconvolved using a model which
does include the seed photon curvature in the comptonised spectrum. 
This dramatically changes the derived boundary layer temperature,
reducing it to $\sim 0.3$ keV. 
Comptonization approximations which do not take into account the seed
photons can give highly misleading results for the boundary layer emission.}
\label{fig:c_cst_bb}
\end{figure*}

Figure \ref{fig:c_cst_bb} 
compares the derived spectra for datasets 
02 and 08 using the black body boundary layer plus comptonised disc
approach. The upper panel shows the fits where the seed photon
curvature is neglected, which is the standard approach in the
literature, while the lower panel shows how the fits change when the seed
photon curvature is included in the models.
It is clear that the blackbody boundary layer temperature changes
dramatically, from $\sim 1$ keV derived from the inapplicable 
{\tt compst} model to $\sim 0.3$ keV, in models with the seed photon compton
curvature ({\tt thcomp}). 

\begin{table*}
\begin{minipage}{180mm}
\caption{GINGA data fit to the Eastern model with comptonization of
the boundary layer including seed photon curvature and full reflection
({\tt diskbb + thcomp + reflection})}
\label{tab:dcref}
\begin{tabular}{ccccccccccc}                \hline

data & $kT_{\rm diskbb}$ keV & $f_{\rm diskbb}$
\footnote{all fluxes are the unabsorbed model, integrated from
0.01--100 keV, in units of $10^{-8}$ ergs s$^{-1}$ cm$^{-2}$} 
& $\Gamma$ &
$kT_{\rm e}$ keV  & $kT_{\rm seed}$ keV & 
$f_{\rm thcomp}^a$ & $\Omega/2\pi$ & $\xi$ & $R_{\rm in}$ &
$\chi^2/\nu$ \\
\hline
01 & $0.53_{-0.20}^{+0.30}$ & 0.41 & $2.30_{-0.11}^{+0.22}$ & 
$3.36_{-0.14}^{+0.15}$ & $0.80_{-0.12}^{+0.24}$ & 1.14 & 
$0.13_{-0.09}^{+0.58}$ & $110_{-110}^{+9000}$ & $38_{-32}^{+\infty}$ &
$8.4/20$\\
02 & $0.51_{-0.13}^{+0.31}$ & 0.43 & $2.32_{-0.08}^{+0.25}$  &
$3.26_{-0.10}^{+0.17}$ & $0.82_{-0.07}^{+0.21}$ & 1.27 &
$0.07_{-0.03}^{+0.57}$ & $670_{-670}^{+18000}$ & $42_{-36}^{+\infty}$
& $9.0/20$ \\
03 & $0.61_{-0.15}^{+0.90}$ & 0.44 & $2.49_{-0.11}^{+0.32}$ &
$3.41_{-0.20}^{+0.50}$  & $0.90_{-0.09}^{+1.22}$ & 1.30 &
$0.08_{-0.04}^{+0.57}$  & $570_{-570}^{+20000}$ & $49_{-43}^{+\infty}$
& $12.0/20$ \\
04 & $0.58_{-0.18}^{+0.99}$ & 0.44 & $2.53_{-0.17}^{+0.25}$ &
$3.31_{-0.26}^{+0.36}$  & $0.87_{-0.13}^{+0.18}$ & 1.26 &
$0.11_{-0.06}^{+0.15}$  & $9300_{-9260}^{+60000}$ &
$49_{-43}^{+\infty}$  & $14.6/20$ \\
05 & $0.41_{-0.13}^{+0.15}$ & 0.68 &
$2.88_{-0.16}^{+0.12}$ & $2.98_{-0.19}^{+0.30}$ & $0.78\pm 0.06$ &
1.20 & $0.21_{-0.06}^{+0.24}$ & $1000_{-900}^{+8000}$ & $40_{-25}^{+240}$ & 
$24.8/20$ \\
06 & $0.39_{-0.14}^{+0.18}$ & 0.74 &
$2.90_{-0.16}^{+0.30}$ & $2.90_{-0.21}^{+0.45}$ &
$0.80^{+0.08}_{-0.05}$ & 1.26 & $0.25_{-0.09}^{+0.10}$ & $240_{-140}^{+1600}$
& $1000_{-985}^{+\infty}$ & $15.6/20$ \\
07 & $0.36_{-0.14}^{+0.18}$ & 0.89 &
$2.77_{-0.14}^{+0.25}$ & $2.77_{-0.14}^{+0.33}$ & $0.80_{-0.05}^{+0.07}$ & 
1.26 & $0.16_{-0.05}^{+0.22}$ & $2100_{-2000}^{+13000}$ &
$90_{-75}^{+\infty}$ & $22.6/20$ \\
08 & $0.35_{-0.20}^{+0.19}$ & 0.82 & $2.69_{-0.15}^{+0.21}$ &
$2.74_{-0.14}^{+0.26}$ & $0.81_{-0.06}^{+0.07}$ & 1.28 &
$0.11_{-0.05}^{+0.18}$ & $1200_{-1110}^{+16000}$ &
$92_{-80}^{+\infty}$  & $12.4/20$ \\ 
09 & $0.34_{-0.23}^{+0.28}$ & 0.64 & $2.54_{-0.11}^{+0.19}$ &
$2.66_{-0.12}^{+0.19}$ & $0.80_{-0.05}^{+0.09}$ & 1.13 &
$0.11_{-0.06}^{+0.11}$ & $1.80^{+4.8}_{-1.74}\times 10^4$ & 
$1000^{+\infty}_{-994}$ & $17.4/20$\\

\end{tabular}
\end{minipage}
\end{table*}

\begin{table*}
\begin{minipage}{180mm}
\caption{GINGA data fit to the Eastern model with comptonization of
the boundary layer including seed photon curvature with approximate reflection
({\tt diskbb + thcomp + gau})}
\label{tab:dcgau}
\begin{tabular}{ccccccccccc}                \hline

data & $kT_{\rm diskbb}$ keV & $f_{\rm diskbb}$
\footnote{all fluxes are the unabsorbed model, integrated from
0.01--100 keV, in units of $10^{-8}$ ergs s$^{-1}$ cm$^{-2}$} 
& $\Gamma$ &
$kT_{\rm e}$ keV &  $kT_{\rm seed}$ keV & $f_{\rm thcomp}^a$ & 
$E_{\rm line}$ keV\footnote{Energy constrained to be between 6.4--7 keV}
& $\sigma$ keV \footnote{Width constrained to be between 0--0.42 keV}
& EW (eV) & $\chi^2/\nu$ \\
\hline
01 & $0.46_{-0.12}^{+0.11}$ & 0.47 & $2.30\pm 0.05$ & 
$3.42_{-0.10}^{+0.11}$ & $0.78_{-0.05}^{+0.06}$ & 1.20 & $6.7\pm 0.3$ & 
$0.42_{-0.41}$ & $50\pm 18$ & 11.0/20\\
02 & $0.46_{-0.09}^{+0.14}$ & 0.51 & $2.35_{-0.04}^{+0.06}$ &
$3.34_{-0.07}^{+0.11}$ & $0.83_{-0.04}^{+0.05}$ & 1.31 & $6.8\pm 0.2$ &
$0.42_{-0.42}$ & $52_{-22}^{+11}$& 12.5/20\\
03 & $0.53_{-0.11}^{+0.13}$ & 0.48 & $2.53_{-0.07}^{+0.08}$ &  
$3.50_{-0.13}^{+0.17}$ & $0.88_{-0.04}^{+0.07}$ & 1.36 &
$6.9^{+0.1}_{-0.3}$ & $0.42_{-0.42}$ & $50_{-20}^{+17}$  & 15.1/20 \\
04 & $0.52_{-0.10}^{+0.15}$ & 0.51 & $2.67_{-0.10}^{+0.12}$ &  
$3.58_{-0.19}^{+0.30}$ & $0.89_{-0.05}^{+0.07}$ & 1.36 & $7.0_{-0.2}$ &
$0.42_{-0.27}$ & $52_{-19}^{+18}$ & 20.2/20\\
05 & $0.45_{-0.07}^{+0.09}$ & 0.73 & $3.30_{-0.14}^{+0.20}$ & 
$3.77_{-0.34}^{+0.62}$ &  $0.86_{-0.03}^{+0.05}$ & 1.24 &
$6.8_{-0.2}^{+0.1}$ & $0.42_{-0.08}$ & $97\pm 20$ & 47.7/20 \\
06 & $0.42_{-0.10}^{+0.12}$ & 0.74 & $3.24_{-0.20}^{+0.25}$ & 
$3.53_{-0.43}^{+0.70}$ & $0.87_{-0.04}^{+0.05}$ & 1.29 & $6.7\pm 0.2$ & 
$0.42_{-0.10}$ & $100_{-23}^{+26}$ & 29.5/20 \\
07 & $0.41_{-0.08}^{+0.11}$ & 0.78 & $3.15_{-0.14}^{+0.22}$ & 
$3.32_{-0.34}^{+0.50}$ & $0.87_{-0.03}^{+0.05}$ & 1.31 & $6.8\pm 0.2$ & 
$0.42_{-0.14}$ & $86_{-25}^{+0.20}$ & 38.8/20 \\
08 & $0.39_{-0.12}^{+0.13}$ & 0.67 & $2.89_{-0.12}^{+0.17}$ & 
$3.00_{-0.18}^{+0.27}$ & $0.86_{-0.04}^{+0.05}$ & 1.30 & $6.8\pm 0.2$ & 
$0.42_{-0.42}$ & $60_{-25}^{+15}$ & 15.8/20\\
09 & $0.37_{-0.13}^{+0.22}$ & 0.59 & $2.72_{-0.09}^{+0.17}$ & 
$2.83_{-0.12}^{+0.22}$ & $0.85_{-0.04}^{+0.06}$ & 1.20 & $7.0_{-0.2}$ & 
$0.42_{-0.42}$ & $50_{-23}^{+16}$ & 21.3/20 \\

\end{tabular}
\end{minipage}
\end{table*}

\begin{table*}
\begin{minipage}{180mm}
\caption{GINGA data fit to the Eastern model with approximate
comptonization of
the boundary layer and approximate reflection
({\tt diskbb + compst + gaussian})}
\label{tab:dcstgau}
\begin{tabular}{cccccccccc}                \hline

data & $kT_{\rm diskbb}$ keV & $f_{\rm diskbb}$
\footnote{all fluxes are the unabsorbed model, integrated from
0.01--100 keV, in units of $10^{-8}$ ergs s$^{-1}$ cm$^{-2}$} 
& $kT_{\rm e}$ (keV) &
$\tau_{\rm e}$ & $f_{\rm compst}^a$ 
& $E_{\rm line}$ keV\footnote{Energy constrained to be between 6.4--7 keV}
& $\sigma$ keV \footnote{Width constrained to be between 0--0.42 keV}
& EW (eV) & $\chi^2/\nu$ \\
\hline
01 & $1.63_{-0.05}^{+0.06}$ & 0.70 & $2.93\pm 0.07$ &
$14.5_{-1.0}^{+1.4}$ & 1.06 & $6.7\pm 0.2$  & $0.42_{-0.42}$ & 
$50_{-20}^{+15}$ & 10.4/21 \\
02 & $1.68\pm 0.04$ & 0.94 & $2.86_{-0.05}^{+0.08}$ & 
$16.6_{-1.5}^{+3.5}$ & 0.82 & $6.7_{-0.2}^{+0.3}$ & $0.42_{-0.42}$  & 
$47_{-14}^{+15}$ & 13.0/21 \\
03 & $1.70\pm 0.04$ & 1.14 & $2.87_{-0.11}^{+0.10}$ &
$17.4_{-2.1}^{+5.7}$ & 0.68 & $6.8_{-0.3}^{+0.2}$ & $0.42_{-0.42}$ & 
$48_{-14}^{+13}$ & 16.9/21 \\
04 & $1.68\pm 0.04$ & 1.24 & $2.82_{-0.17}^{+0.14}$ & 
$18.3_{-3.0}^{+\infty}$ & 0.56 & $6.9_{-0.2}^{+0.1}$ & $0.42_{-0.28}$ & 
$53_{-17}^{+18}$ & 21.9/21 \\
05 & $1.51\pm 0.02$ & 1.50 & $2.44_{-0.05}^{+0.22}$ &
$200_{-185}^{+\infty}$ & 0.16 & $6.7_{-0.2}^{+0.1}$ & $0.42_{-0.06}$ &
$105\pm 20$ & 50.6/21 \\
06 & $1.58_{-0.05}^{+0.02}$ & 1.53 & $2.55_{-0.10}^{+0.11}$ & 
$200_{-180}^{+\infty}$ & 0.14 & $6.5_{-0.1}^{+0.2}$ & $0.42_{-0.10}$ & 
$95\pm 20$ & 41.0/21  \\
07 & $1.61_{-0.02}^{+0.03}$ & 1.55 & $2.58_{-0.09}^{+0.10}$ &
$200_{-177}^{+\infty}$ & 0.14 & $6.6\pm 0.2$ & $0.42_{-0.14}$ & 
$73_{-12}^{+17}$ & 59.7/21 \\
08 & $1.66_{-0.02}^{+0.03}$ & 1.49 & $2.59\pm 0.09$ & 
$200_{-175}^{+\infty}$ & 0.14 & $6.5_{-0.1}^{+0.4}$ & $0.42_{-0.42}$ & 
$50_{-20}^{+18}$ & 47.6/21 \\
09 & $1.72_{-0.03}^{+0.02}$ & 1.35 & $2.63_{-0.08}^{+0.10}$ &
$200_{-175}^{+\infty}$ & 0.13 & $6.7\pm 0.3$ & $0^{+0.42}$ & 
$22_{-12}^{+14}$ & 60.8/21 \\

\end{tabular}
\end{minipage}
\end{table*}

\begin{table*}
\begin{minipage}{180mm}
\caption{GINGA data fit to the Western model with approximate 
comptonization of the disc and approximate reflection
({\tt bbody +  compst + gau}).}
\label{tab:cstbbgau}
\begin{tabular}{cccccccccc}                \hline

data & $kT_{\rm bb}$ (keV) & $f_{\rm bb}$ 
\footnote{all fluxes are the unabsorbed model, integrated from
0.01--100 keV, in units of $10^{-8}$ ergs s$^{-1}$ cm$^{-2}$} 
& $kT_e$ (keV) &
$\tau$ & $f_{\rm compst}^a$ 
& $E_{\rm line}$ keV\footnote{Energy constrained to be between 6.4--7 keV}
& $\sigma$ keV \footnote{Width constrained to be between 0--0.42 keV}
& EW (eV) & 
$\chi^2/\nu$ \\
\hline
01 & $1.16\pm 0.03$ & 0.26 & $2.95\pm 0.05$ & $12.0\pm 0.3$ & 
1.86 & $6.8\pm 0.2$ & $0.2\pm 0.2$ & $35\pm 17$ & 7.5/21\\
02 & $1.18\pm 0.03$ & 0.34 & $2.87\pm 0.04$ & $12.4\pm 0.3$ & 
1.92 & $6.9_{-0.3}^{+0.1}$ & $0.3^{+0.1}_{-0.3}$ & $30_{-16}^{+20}$ & 9.7/21\\
03 & $1.19\pm 0.03$ & 0.41 & $2.84^{+0.06}_{-0.04}$ & $12.2\pm 0.3$ &
2.02 & $7.0_{-0.3}$ & $0.4_{-0.4}$ & $34\pm 20$ & 17.0/21\\
04 & $1.16\pm 0.03$ & 0.44 & $2.78^{+0.08}_{-0.06}$ & $12.1\pm 0.4$ & 
2.06 & $7.0_{-0.2}$ & $0.4_{-0.2}$ & $44_{-22}^{+16}$ & 19.4/21\\
05 & $1.09\pm 0.02$ & 0.51 & $2.55\pm 0.07$ & $11.1\pm 0.4$ & 
2.45 & $6.8\pm 0.1$ & $0.4_{-0.1}$ & $80_{-10}^{+20}$ & 32.2/21\\
06 & $1.09\pm 0.03$ & 0.55 & $2.46\pm 0.10$ & $11.9\pm 0.6$ & 
2.12 & $6.7\pm 0.2$ & $0.4_{-0.1}$ & $90\pm 25$ & 21.2/21\\
07 & $1.10\pm 0.03$ & 0.55 & $2.44\pm 0.08$ & $12.2\pm 0.6$ & 
2.01 & $6.9^{+0.1}_{-0.2}$ & $0.4_{-0.2}$ & $70\pm 20$ & 30.9/21\\
08 & $1.10\pm 0.03$ & 0.52 & $2.38\pm 0.06$ & $13.1\pm 0.6$ & 
1.73 & $6.8\pm 0.2$ & $0^{+0.4}$ & $38_{-16}^{+30}$& 11.9/21\\
09 & $1.09\pm 0.03$ & 0.46 & $2.34\pm 0.05$ & $14.0\pm 0.6$ & 
1.44 & $7.0_{-0.2}$  & $0.4_{-0.4}$ & $38_{-23}^{+18}$ & 21.7/21\\

\end{tabular}
\end{minipage}
\end{table*}

\begin{table*}
\begin{minipage}{180mm}
\caption{GINGA data fit to the Western model with full
comptonization of the disc and approximate reflection
({\tt bbody +  thcomp + gau}).}
\label{tab:cbbgau}
\begin{tabular}{ccccccccccc}                \hline

data & $kT_{\rm bb}$ (keV) & $f_{\rm bb}$ 
\footnote{all fluxes are the unabsorbed model, integrated from
0.01--100 keV, in units of $10^{-8}$ ergs s$^{-1}$ cm$^{-2}$} 
& $\Gamma$ (keV) &
$kT_e$ & $kT_{seed}$ & $f_{thcomp}^a$ 
& $E_{\rm line}$ keV\footnote{Energy constrained to be between 6.4--7 keV}
& $\sigma$ keV \footnote{Width constrained to be between 0--0.42 keV}
& EW (eV) & 
$\chi^2/\nu$ \\
\hline
01 & $0.36_{-0.06}^{+0.07}$ & 0.24 & $2.29_{-0.04}^{+0.05}$ &
$3.42_{-0.09}^{+0.10}$ & $0.78\pm 0.05$ & 1.19 & $6.8\pm 0.2$ &
$0.42_{-0.42}$ & $50_{-20}^{+18}$ & 10.5/20\\
02 & $0.38\pm 0.06$ & 0.25 & $2.36\pm 0.05$ & $3.35_{-0.08}^{+0.10}$ &
$0.83\pm 0.04$ & 1.31 & $6.8\pm 0.2$ & $0.42_{-0.42}$ &
$50_{-20}^{+12}$ & 11.9/20\\ 
03 & $0.41\pm 0.06$ & 0.26 & $2.52\pm 0.07$ & $3.49_{-0.12}^{+0.17}$ &
$0.88\pm 0.04$ & 1.37 & $6.9_{-0.2}^{+0.1}$ & $0.42_{-0.42}$ &
$48_{-20}^{+18}$ & 14.4/20\\
04 & $0.41\pm 0.06$ & 0.27 & $2.67\pm 0.10$ & $3.57_{-0.20}^{+0.28}$ &
$0.89\pm 0.04$ & 1.37 & $7.0_{-0.2}$ & $0.42_{-0.42}$ &
$51_{-20}^{+18}$ & 19.5/20\\
05 & $0.37\pm 0.05$ & 0.36 & $3.31_{-0.14}^{+0.18}$ & $3.79_{-0.34}^{+0.61}$
& $0.86\pm 0.03$ & 1.25 & $6.8_{-0.2}^{+0.1}$ & $0.42_{-0.10}$ & $95\pm 20$
& 46.7 /20\\ 
06 & $0.35\pm 0.07$ & 0.35 & $3.25_{-0.20}^{+0.23}$ & $3.54_{-0.42}^{+0.69}$
& $0.87\pm 0.04$  & 1.29 & $6.7\pm 0.2$ & $0.42_{-0.10}$ & $100\pm 25$
& 29.1/20\\
07 & $0.34\pm 0.06$ & 0.36 & $3.15_{-0.13}^{+0.21}$ &
$3.32_{-0.16}^{+0.47}$ & $0.87\pm 0.04$  & 1.31 &  $6.8_{-0.1}^{+0.2}$ &
$0.42_{-0.14}$ & $85_{-25}^{+10}$ & 38.3/20\\
08 & $0.32\pm 0.08$ & 0.33 & $2.89_{-0.10}^{+0.16}$ & $2.99_{-0.14}^{+0.25}$
& $0.86\pm 0.04$ & 1.30& $6.8\pm 0.2$ & $0.42_{-0.42}$ &
$64_{-30}^{+16}$ & 15.7/20\\
09 & $0.34_{-0.12}^{+0.08}$ & 0.23 & $2.74_{-0.11}^{+0.13}$ &
$2.86_{-0.15}^{+0.16}$  & $0.86\pm 0.04$ & 1.43 & $7.0_{-0.2}$ &
$0.42_{-0.42}$ & $45\pm 21$ & 21.2/20\\

\end{tabular}
\end{minipage}
\end{table*}

\section{DISCUSSION AND CONCLUSIONS}

The spectrum of Cyg X--2 has complex curvature which makes spectral
fitting highly non--unique. It is generally agreed that the X--ray
spectrum must consist of at least two components, the disc and the
boundary layer, and that at least one of these (though more probably
both) is comptonised. In addition, there are spectral features
corresponding to ionised iron K, which are often modelled as a broad
Gaussian line. We have shown that the derived spectral parameters are
crucially dependent on how the comptonised spectrum is modelled.  The
often used {\tt compst} or cutoff power law models
are physically inconsistent since the
seed photons for the compton scattering {\em must} be in the observed
X--ray range, whether they come from the disc or the boundary layer.
The lack of curvature in {\tt compst} or cutoff power laws
at energies below the electron
temperature then forces a blackbody at $\sim 1$ keV into the fit. This
is purely an artifact of incorrect modelling of the comptonised
spectrum and has {\it no} physical relation to the boundary layer or neutron
star surface. 

When the comptonised spectrum is modelled including this curvature
(using {\tt thcomp} or {\tt comptt}) then the spectrum is adequately
fit either by assuming that the boundary layer is comptonised, and the
softer emission is from the accretion disc ({\tt diskbb+thcomp+gau}
$\chi^2_\nu=211.9/180$) or that the disc spectrum is comptonised, and
the boundary layer is a blackbody ({\tt bb+thcomp+gau}
$\chi^2_\nu=207.3/180$). The GINGA data cannot distinguish between
these models statistically. However, there are some theoretical
constraints on the ratio of the disc and boundary layer fluxes. In the
`Eastern model', where the boundary layer is comptonised, we derive a
boundary layer luminosity which is about twice that of the luminosity
of the disc. This is as expected in the general relativistic models of
the gravitational potential if the neutron star spin is not close to
the Keplarian breakup frequency (Sunyaev \& Shakura 1986). By
contrast, in the `Western model' where the disc is strongly
comptonised, the ratio of the 
boundary layer blackbody luminosity to that from the disc is
$\sim 0.2-0.25$, implying that the neutron star is
significantly spun up. Detailed calculations indicate that to
get a ratio of boundary layer to disc flux this low
requires that the spin frequency of the neutron star is $\ge 1$ kHz
(Sibgatullin \& Sunyaev 2000).  This seems excessive: all known
millisecond pulsars (the descendants of the LMXB neutron stars)
have spin frequencies $\le 700$ Hz (e.g.  Kramer
et al., 1998). 

Thus it seems likely that the `Eastern' model (Mitsuda et al., 1984;
1989) is correct and that the boundary layer dominates the hard
spectrum, while the disc dominates at softer energies. A standard 
objection to this approach, that in general
the derived disc parameters require
an emitting area which is far too small to be a true disc (e.g. Church
\& Balucinska--Church 2001), is again purely an artifact of incorrect
modelling of the curvature due to the seed photons in the 
comptonised continuum. 

Finally, it is clear that the residual spectral features at iron
are considerably better described by a strongly ionised 
reflected spectrum than by a broad Gaussian line.

\end{document}